\documentclass[a4paper,onecolumn, 10pt]{article} 
 
\usepackage{amsmath,amssymb}
\usepackage[utf8]{inputenc}
\usepackage{authblk}

\usepackage{abstract}
\usepackage[a4paper, top=25mm, bottom=25mm]{geometry}
\usepackage[colorlinks,allcolors=black,bookmarks=false,hypertexnames=true]{hyperref} 
\usepackage[backend=biber, style=phys, sorting=none, doi=false,isbn=false,url=false, minbibnames=1, maxbibnames=5]{biblatex}
\AtEveryBibitem{\clearfield{month}}
\AtEveryBibitem{\clearfield{day}}
\AtEveryBibitem{\clearfield{issue}}

\usepackage{textcomp}

\usepackage{graphicx}
\usepackage[margin=10pt,font=small,labelfont=bf,format=plain]{caption} 

\title{Coherent Light Control of a Metastable Hidden Phase} 

\author[1]{J. Maklar}
\author[1]{S. Dong}
\author[1]{J. Sarkar}
\author[2,3]{Y.A. Gerasimenko}
\author[1]{T. Pincelli}
\author[1,$\ast$]{S. Beaulieu}
\author[4]{P.S. Kirchmann}
\author[4]{J.A. Sobota}
\author[4,5,$\dag$]{S.-L. Yang}
\author[4,5]{D. Leuenberger}
\author[4,$\ddag$]{R.G. Moore}
\author[4,5]{Z.-X. Shen}
\author[1]{M. Wolf}
\author[2,3]{D. Mihailovic}
\author[1,6]{R. Ernstorfer}
\author[1]{L. Rettig}

\affil[1]{Fritz-Haber-Institut der Max-Planck-Gesellschaft, Faradayweg 4-6, D-14195 Berlin, Germany}
\affil[2]{Dept. of Complex Matter, Jožef Stefan Institute, Jamova 39, SI-1000 Ljubljana, Slovenia}
\affil[3]{Center of Excellence on Nanoscience and Nanotechnology – Nanocenter (CENN Nanocenter), Jamova 39, SI-1000 Ljubljana, Slovenia}
\affil[4]{Stanford Institute for Materials and Energy Sciences, SLAC National Accelerator Laboratory, 2575 Sand Hill Road, Menlo Park, California 94025, USA}
\affil[5]{Geballe Laboratory for Advanced Materials, Departments of Physics and Applied Physics, Stanford University, Stanford, California 94305, USA}
\affil[6]{Institut für Optik und Atomare Physik, Technische Universität Berlin, Straße des 17. Juni 135, 10623 Berlin, Germany}
\affil[$\ast$]{Present address: Université de Bordeaux - CNRS - CEA, CELIA, UMR5107, F33405, Talence, France}
\affil[$\dag$]{Present address: University of Chicago, Pritzker School of Molecular Engineering, Chicago, IL 60637, USA}
\affil[$\ddag$]{Present address: Materials Science and Technology Division, Oak Ridge National Laboratory, Oak Ridge, TN 37831, USA}
\date{\today}

\begin{document}

    \maketitle

    \begin{abstract}
    Metastable phases present a promising route to expand the functionality of complex materials. Of particular interest are light-induced metastable phases that are inaccessible under equilibrium conditions, as they often host new, emergent properties switchable on ultrafast timescales. However, the processes governing the trajectories to such hidden phases remain largely unexplored. Here, using time- and angle-resolved photoemission spectroscopy, we investigate the ultrafast dynamics of the formation of a hidden quantum state in the layered dichalcogenide 1$T$-TaS$_2$ upon photoexcitation. Our results reveal the nonthermal character of the transition governed by a collective charge-density-wave excitation. Utilizing a double-pulse excitation of the structural mode, we show vibrational coherent control of the phase-transition efficiency. Our demonstration of exceptional control, switching speed, and stability of the hidden phase are key for device applications.
    \end{abstract}
    
Controlling material properties on demand remains a long-standing challenge in condensed-matter physics. Inspired by femtochemistry, the illumination by ultrashort light pulses poses a promising route to understanding and actively manipulating the macroscopic properties of quantum materials. Multiple optical control pathways have been established based on a transient modification of the free-energy landscape, ultrafast heating, and Floquet engineering~\cite{delaTorre2021_RMP_nonthermal_pathways}. These control schemes allow, for example, triggering electronic phase transitions on femto- to picosecond timescales~\cite{Smallwood2012_Science} and steering intricate structural transitions by excitation of specific lattice modes~\cite{Horstmann2020Jul}. Remarkably, in materials with multiple competing orders, ultrashort light pulses can also induce 
metastable states that feature properties inaccessible under thermodynamic equilibrium conditions, so-called hidden phases. As hidden phases expand emergent functionalities of complex solids, are often switchable on ultrafast timescales, and provide insights into fundamental interactions and phase competition, they are of high interest from a scientific and an application perspective. Prominent examples include optical switching to hidden metallic states in strongly correlated materials~\cite{Morrison2014_Science_VO2,Zhang2016_natmater_manganite} and possible photoinduced superconductivity at temperatures far above $T_\mathrm{c}$~\cite{Budden2021May}. Yet, the microscopic processes governing the transition to metastable states often lack a clear understanding, resulting in unspecific empirical switching protocols with non-optimal efficiencies and limited control over stability. Furthermore, most hidden phases persist only transiently on a ps timescale, precluding any practical application in switchable devices.

\begin{figure}[!htb]
\centering
\includegraphics[width=\textwidth]{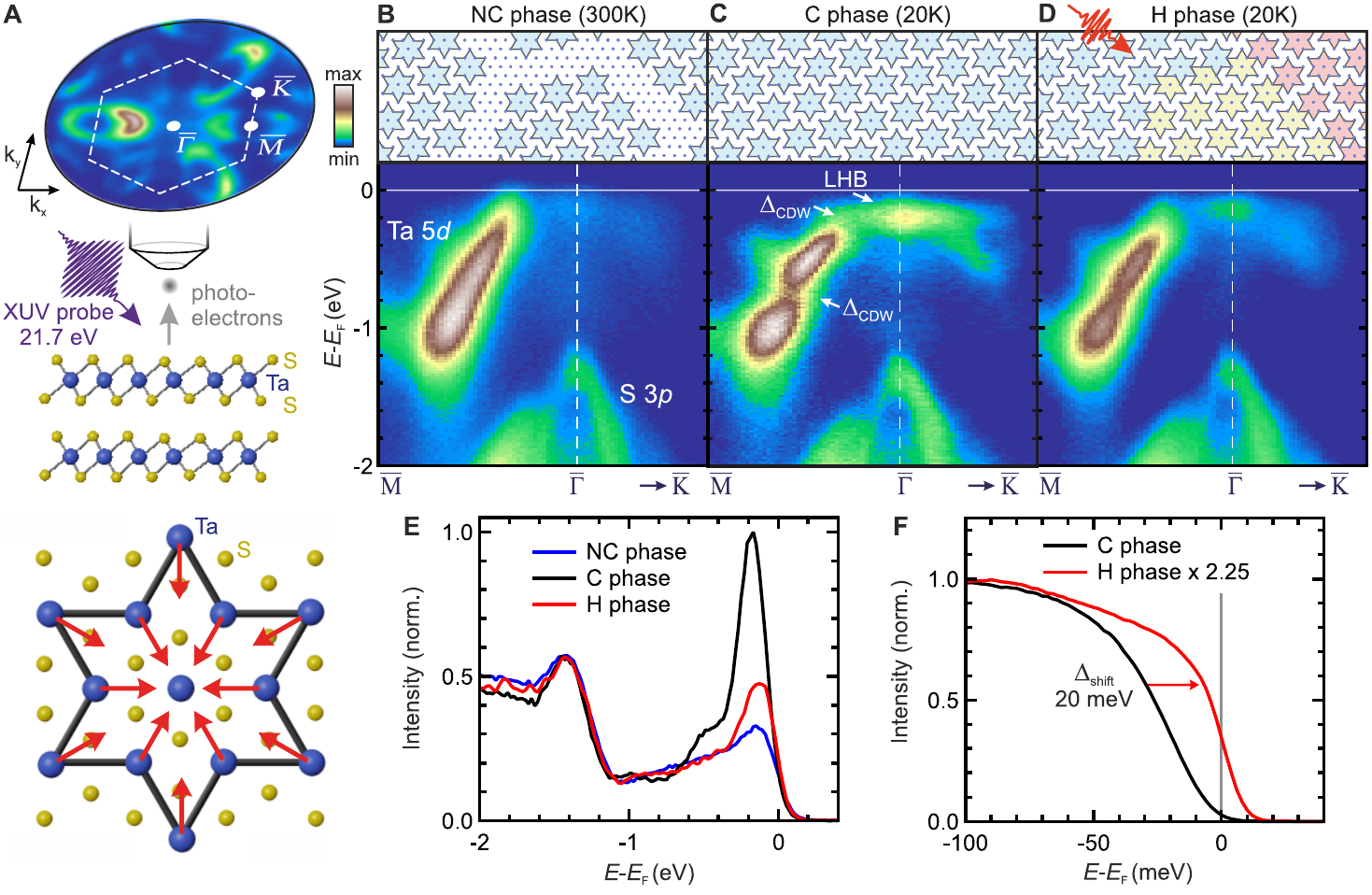}
\caption{\textbf{Band-structure mapping.} (\textbf{A}) Top: Schematic of the static XUV ARPES measurements and the layered 1$T$-TaS$_2$ sample. An exemplary isoenergy surface of the NC phase ($E-E_\mathrm{F}=-0.5$~eV, momentum radius k$_{\parallel}\sim2$~\AA$^{-1}$) with hexagonal surface Brillouin zone and high symmetry points is shown. Bottom: Illustration of the star-shaped polaron cluster. (\textbf{B} to \textbf{D}) Top: Schematic intralayer polaron order of the NC, C and H phases (domains indicated by different colors). Bottom: ARPES energy-momentum cuts along the high-symmetry directions of the respective phases. Switching to the H phase was achieved by a single optical pulse at $h\nu=1.55$~eV at an absorbed fluence $> F_\mathrm{crit} \sim 0.5$~mJ$\cdot$cm$^{-2}$. (\textbf{E}) EDCs of the examined phases at $\overline{\Gamma}$. (\textbf{F}) EDCs of C and H phase at $\overline{\Gamma}$ near $E_\mathrm{F}$ (gray line), obtained with a laser-ARPES setup with high energy resolution (see methods).}
\label{fig:static}
\end{figure}

A particularly interesting metastable state is the metallic hidden (H) phase of 1$T$-TaS$_2$, as it can be reversibly induced by optical or electrical pulses and is accompanied by an insulator-to-metal transition with an orders-of-magnitude drop in resistivity~\cite{Stojchevska2014Apr,Vaskivskyi2015Jul,Ma2016_originofmott,Cho2016_electric_switch_interlayer_stacking,gerasimenko_2019_Hphase_order,Stahl2020_NatCom,Ravnik2021Apr,Ravnik2018_PRB_realtime_H,Gao2021_arXiv_dynamics_Hphase}. Additionally, the H phase is exceptionally long-lived and thus represents a promising platform for device functionality~\cite{Mihailovic2021_APL_memory_devices}. In equilibrium, the layered transition-metal dichalcogenide 1$T$-TaS$_2$ features multiple competing ground states resulting from the interplay of Coulomb repulsion, Fermi-surface nesting, and interlayer interaction. At the heart of the various orderings lies the formation of star-shaped polaron clusters, i.e., clusters of ions that are inwards radially displaced toward an unbound charge localized at the cluster centers (Fig.~\ref{fig:static}A). The polaron formation gives rise to a metallic nearly commensurate (NC) phase below 350~K that consists of patches of ordered polaron clusters separated by smooth periodic domain walls (Fig.~\ref{fig:static}B). Cooling below 180~K leads to a hysteretic first-order transition to a Mott-insulating commensurate (C) phase with long-range intralayer polaron order~\cite{Thomson1994_PRB_STM_TaS2} (Fig.~\ref{fig:static}C), also discussed in terms of a charge density wave (CDW). Additionally, in the C phase, an interlayer alignment of polarons leads to a unit cell doubling along the out-of-plane direction, referred to as interlayer stacking order, which is suspected to strongly affect the low-energy electronic states~\cite{Ma2016_originofmott,Butler2020May,Ritschel2015Apr,Lee_2019_PRL_DFT_bandinsulator, Nicholson2022_arXiv}.

Optical or electrical excitation of the C ground state triggers a transition to the metallic metastable H phase, which consists of microscopic commensurate domains intersected by a large number of irregular domain walls that induce an abrupt CDW phase slip between neighboring domains (Fig.~\ref{fig:static}D). While the atomic structure of the H phase has been studied extensively~\cite{Ma2016_originofmott,Cho2016_electric_switch_interlayer_stacking,gerasimenko_2019_Hphase_order,Stahl2020_NatCom}, the origin of the metallization is still under debate~\cite{Stojchevska2014Apr,Cho2016_electric_switch_interlayer_stacking,Butler2020May,Stahl2020_NatCom} as measurements of the electronic band structure have been missing. Furthermore, even though an electron-hole imbalance following the excitation has been suggested as driver of the transition to the H phase~\cite{Stojchevska2014Apr}, a microscopic understanding of the switching mechanism -- imperative for improving the switching energy and speed of memory devices based on metastable configurations~\cite{Mihailovic2021_APL_memory_devices} -- has remained elusive. Particularly the role of coherences, which have been exploited to boost the switching efficiency in other systems~\cite{Horstmann2020Jul}, has remained largely unexplored.

Here, we use time- and angle-resolved photoemission spectroscopy (trARPES) to probe the electronic band structure and formation dynamics of the metastable H quantum state in bulk 1$T$-TaS$_2$ upon near-infrared femtosecond optical excitation. Employing a combination of XUV and UV trARPES, we map out the band structure of the H phase within the full first Brillouin zone and at high energy resolution. We observe a global collapse of the Mott-insulating state, providing further hints for the critical role of the interlayer stacking order. Tracking the electron dynamics during the photoinduced phase transition from C ground state to H phase reveals a surprisingly fast timescale of a few hundred fs, suggesting a highly efficient nonthermal switching pathway governed by a collective CDW excitation. Using a two-pump-pulse excitation scheme, we demonstrate vibrational coherent control of the transition to the H phase, evident from pump-pump-delay-dependent oscillations of the switching efficiency at the CDW amplitude mode frequency. Based on the observed band-structure dynamics, we discuss an ultrafast microscopic switching pathway from C to H phase. Our study demonstrates exceptional control over the properties of complex solids by utilizing tailored optical multi-pulse excitations of ordered states and highlights a promising route in the search for new hidden states.

\section*{Results}
\textbf{Band-structure mapping.} First, we map out the static band structure along the high-symmetry directions for the different phases (Fig.~\ref{fig:static}). In the NC phase at room temperature, Ta 5$d$ states give rise to a metallic low-energy band that reaches up to the Fermi level ($E_\mathrm{F}$) at $\overline{\Gamma}$. Upon cooling, the gapped C phase emerges, as the polaron clusters become commensurate with the atomic lattice. This periodic charge- and lattice distortion leads to a splitting of the Ta 5$d$ band into subbands due to the formation of CDW energy gaps $\Delta_\mathrm{CDW}$, with $d$ electrons of the 12 outer Ta atoms of the star-shaped clusters occupying the subbands up to $E-E_\mathrm{F}\sim-0.4$~eV. The remaining 13th electron remains localized on the central Ta atom due to strong on-site Coulomb repulsion, giving rise to a narrow lower Hubbard band (LHB) at -0.2~eV and the opening of a small Mott gap at $E_\mathrm{F}$, in agreement with previous observations~\cite{Perfetti2005_PRB_mapping}. Strikingly, switching from the C ground state to the persistent H phase at 20~K by a single optical pulse induces a drastic modification of the low-energy states. Most notably, the LHB intensity is strongly suppressed, as evinced by the energy distribution curves (EDCs) at $\overline{\Gamma}$ (Fig.~\ref{fig:static}E), accompanied by a suppression of the Mott gap (Fig.~\ref{fig:static}F and Supplementary Fig.~\ref{fig:S_6eV}). While, strictly speaking, the low-energy band in the H phase is a strongly-correlated metallic band, we refer to it as LHB across all phases.

Since ARPES is only sensitive to short-range order and averages over a macroscopic surface area due to the \textmu m-sized probe beam, the drastic spectral change upon switching from C to H phase is remarkable, as both phases consist predominantly of commensurately ordered polaron clusters, with only a fraction of the H phase consisting of domain walls (Supplementary Fig.~\ref{fig:S_stm_images}). However, recent studies demonstrated a strong influence of the interlayer polaron stacking order on the low-energy electronic states~\cite{Ma2016_originofmott,Butler2020May,Ritschel2015Apr,Lee_2019_PRL_DFT_bandinsulator, Nicholson2022_arXiv}. As Ta 5$d_{z^2}$ orbitals with strong out-of-plane character are located at the cluster centers, the interlayer orbital overlap is highly sensitive to the interlayer polaron alignment. While, in the C ground state, two neighboring layers dimerize, switching to the H phase induces a multitude of irregular domain wall networks in each layer which shift the central cluster atoms by one or more atomic lattice vectors in different
directions~\cite{gerasimenko_2019_Hphase_order}. These irregular phase shifts of the intralayer polaron order within each layer induce disorder along the out-of-plane direction, breaking long-range order of the interlayer dimerization~\cite{Stahl2020_NatCom} and thus reducing interlayer hopping. This external restacking may induce metallization, as the subtle balance of on-site Coulomb repulsion, intra- and interlayer hopping is perturbed, quenching the fragile Mott state~\cite{Ma2016_originofmott,Butler2020May}. Our measurements directly support this scenario, as we observe suppression of the LHB and closing of the Mott gap despite only minimal changes in the intralayer ordering. The metallization of a Mott insulator upon reducing interlayer hopping is consistent with a simple bilayer Hubbard model~\cite{Fuhrmann2006}, although a more realistic theoretical treatment is required to capture the respective phases of TaS$_2$.

\begin{figure}[!ht]
\centering
\includegraphics[width=\textwidth]{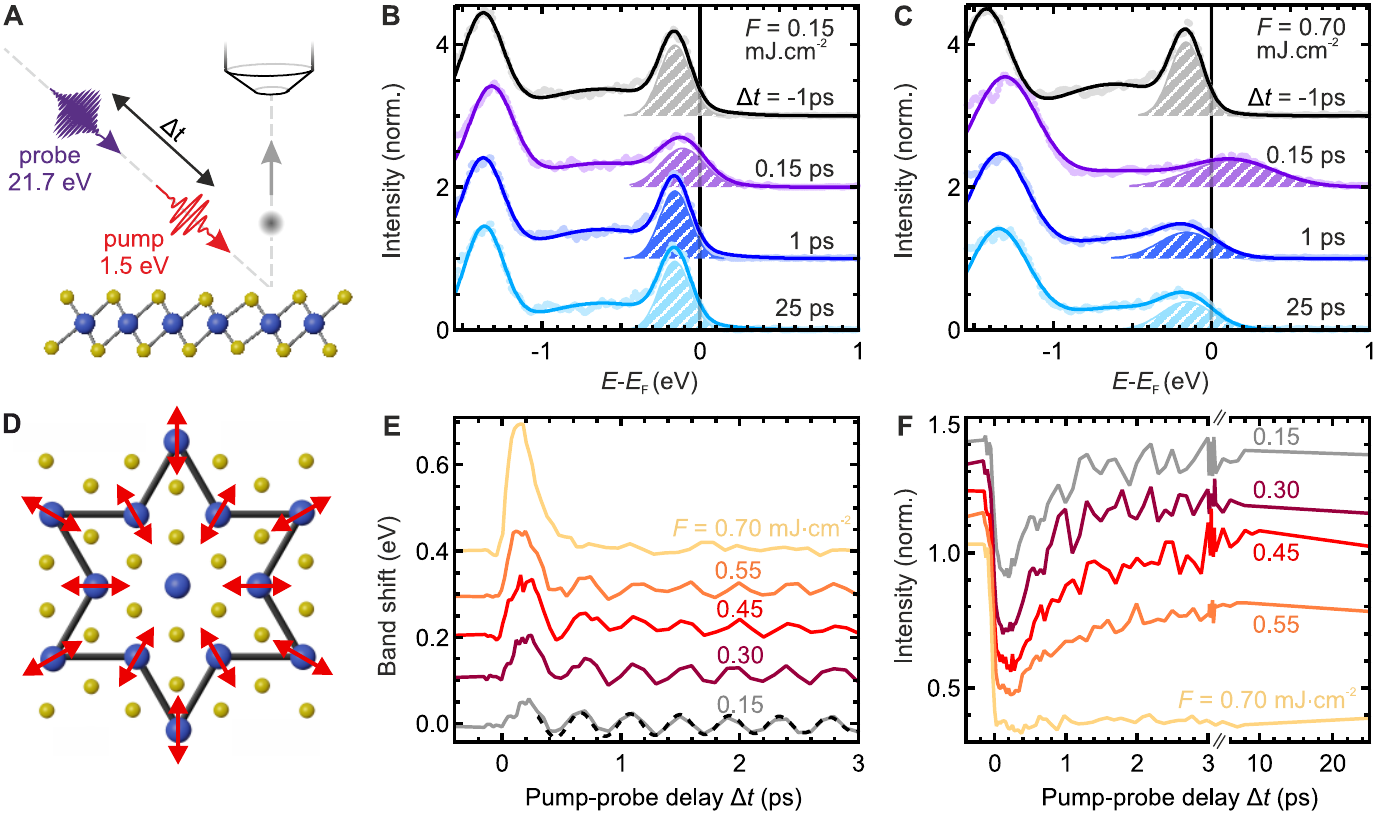}
\caption{\textbf{Dynamical transition to H phase.} (\textbf{A}) Sketch of the trARPES experiment with pump-probe delay $\Delta t$. (\textbf{B}) EDCs (dots) at $\overline{\Gamma}$ for selected delays after excitation at fluences below and (\textbf{C}) above the threshold fluence of the H phase ($T=160$~K, rep.~rate $=10$~kHz). The solid lines mark best fits using 3 Gaussian peaks as described in Supplementary Sec.~\ref{sec:fitting}. The shaded areas indicate the Gaussian peak capturing the LHB. (\textbf{D}) Illustration of the collective oscillation of the star-shaped lattice distortion (CDW AM). (\textbf{E}) Band shift of the LHB versus $\Delta t$ for various fluences, extracted from the corresponding Gaussian center position. The dashed black line marks a sinusoidal fit ($f=2.4$~THz). (\textbf{F}) Peak intensity of the LHB versus $\Delta t$ for various fluences, extracted from the Gaussian fits. Curves are vertically offset for clarity.}
\label{fig:pump_probe}
\end{figure}

\textbf{Dynamical phase transition.} Having established the spectral fingerprints of the H phase, we next investigate the dynamical transition from C to H phase. Stroboscopic pump-probe trARPES measurements of the electronic switching dynamics (Fig.~\ref{fig:pump_probe}A) require relaxation to the C ground state between successive pump-probe cycles -- not feasible at very low temperatures due to the exceptionally long lifetime of the H phase~\cite{Vaskivskyi2015Jul}. Thus, we increase the sample temperature to 160~K, which reduces the lifetime to $<100$~\textmu s and allows us to study the transition of the unperturbed C ground state to the H phase in a pump-probe experiment at a repetition rate of 10~kHz. We focus on the dynamics of the LHB at $\overline{\Gamma}$ for various pump-laser fluences ranging from the weak-response regime to above the switching threshold (see exemplary EDCs in Fig.~\ref{fig:pump_probe}, B and C). Using Gaussian fits of the EDCs (see Supplementary Sec.~\ref{sec:fitting}), we extract the dynamical evolution of the LHB peak intensity, the key feature to identify the H phase, as well as the transient shift of the LHB energy position.

For an absorbed fluence of 0.15~mJ$\cdot$cm$^{-2}$, far below the switching threshold, optical excitation launches a collective, coherent oscillation of the periodic charge- and lattice distortion, termed amplitude mode (AM), corresponding to an expansion-contraction motion of the polaron clusters (Fig.~\ref{fig:pump_probe}D) at 2.4~THz~\cite{demsar2002}. Due to strong electron-phonon coupling, this causes a modulation of the LHB energy at the frequency of the AM (grey curve in Fig.~\ref{fig:pump_probe}E). While several bands across the full Brillouin zone are transiently renormalized by the CDW modulation, the energetic position of the LHB at $\overline{\Gamma}$ couples most strongly to the CDW, and can thus be used as a measure of the CDW lattice displacement~\cite{perfetti2006,hellmann2012time}. Additionally, within 50~fs after the excitation, the LHB intensity is reduced by $\sim$40\%, resulting from an initial quench of the Mott state by transient heating and depopulation of the LHB by electron-hole excitations into higher-lying states~\cite{Ligges2018_doublon}, followed by a full recovery within a few ps (grey curve in Fig.~\ref{fig:pump_probe}F).

With increasing fluence, the initial band shift grows while the AM oscillations become less pronounced, and the LHB intensity only partially recovers within the examined time range. Finally, for the highest applied fluence of 0.7~mJ$\cdot$cm$^{-2}$, the system undergoes a transition from C to H phase, evidenced by a strong persistent quench of the LHB intensity that shows no onset of recovery within tens of ps. Concomitantly, within 250~fs, the LHB undergoes a colossal transient metallization, shifting upwards by $\sim$300~meV, far above the equilibrium $E_\mathrm{F}$, followed by a damped exponential recovery. This dramatic shift indicates a pronounced cooperative lattice motion counteracting the CDW distortion, and, as the band shifts significantly above the band position of the metallic high-temperature phases~\cite{sohrt2014fast}, hints toward an overshoot of the displacement beyond the symmetric undistorted state, which has also been observed in the related compound 1$T$-TaSe$_2$~\cite{Zhang2022_CDWinversion}. The absence of AM modulations further supports the scenario of a strong disruption of the CDW order during the phase transition. Within $\sim$800~fs, the LHB position relaxes to the new quasi-equilibrium position and the transition is completed, in general agreement with switching times reported by optical studies~\cite{Ravnik2018_PRB_realtime_H,Ravnik2021Apr,Gao2021_arXiv_dynamics_Hphase}.

Since the transition from C to H phase involves a major reorganization of the lattice, the observed switching time is surprisingly short -- far below the timescale of electron-lattice equilibration~\cite{Mihailovic2021_APL_memory_devices} -- suggesting a highly efficient nonthermal collective transition pathway. The question naturally arises which mode facilitates such a coupled electronic and structural transition. The AM appears to be the most likely driver of such a concerted transition, as it connects electronic and structural orders and evolves, with increasing fluence, from an oscillatory motion into an unusually large band shift associated with a significant cooperative lattice response.

\begin{figure}[!ht]
\centering
\includegraphics[width=\textwidth]{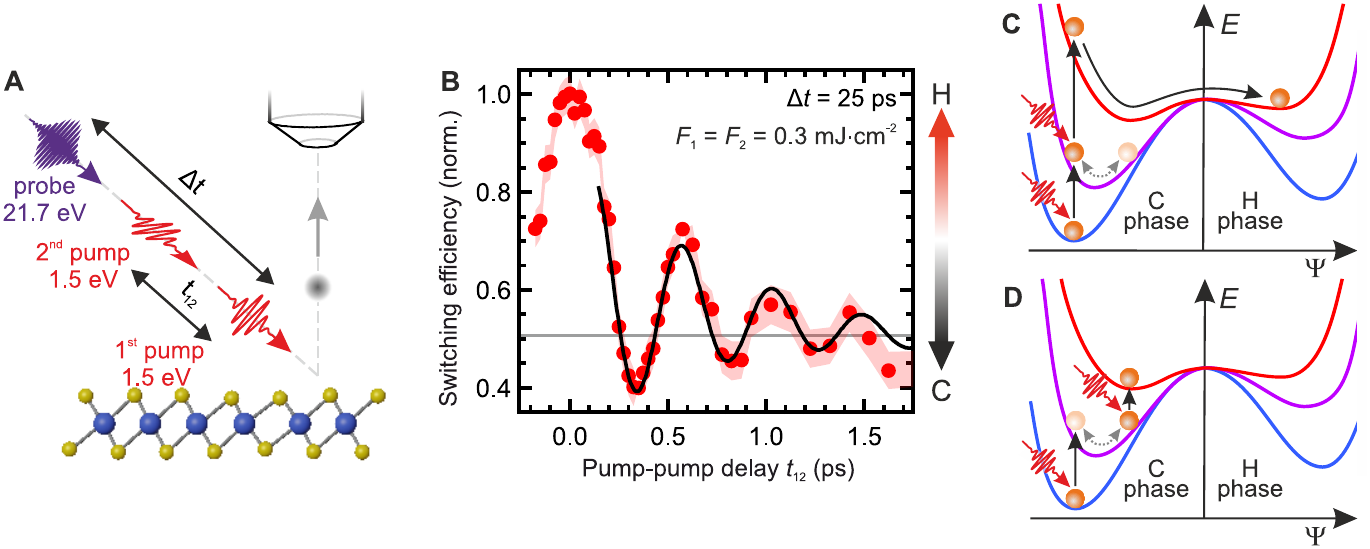}
\caption{\textbf{Coherent control of the transition to the H phase.} (\textbf{A}) Schematic of the trARPES experiment with two successive optical excitation pulses with pump-pump delay $t_{12}$ and pump-probe delay $\Delta t$. To avoid interference artifacts during the cross-correlation, the pump pulses are linearly cross-polarized. (\textbf{B}) Switching efficiency (red dots) versus $t_{12}$, derived from the LHB intensity, with sinusoidal fit (black curve, $f=2.2$~THz). An efficiency of 1 corresponds to the H phase, derived from the LHB intensity at $t_{12}$=0~ps, while an efficiency of 0 corresponds to the LHB intensity of the unperturbed C ground state at 160~K (see Supplementary Sec.~\ref{sec:coherent_switching} for details of the parametrization). The solid gray line indicates the asymptotic behavior for large pump-pump delays. The red shaded area indicates an uncertainty of one standard deviation derived from the Gaussian fits. (\textbf{C}) Schematic energy surfaces in equilibrium (blue) and after excitation by a first (purple) and second (red) optical pulse. The system's order is parametrized by the order parameter $\Psi$ along the AM coordinate. Depending on the pump-pump delay, the excitation leads to an efficient pathway to the H phase or (\textbf{D}) to a halt of the dynamics, suppressing the transition.}
\label{fig:coherent_control}
\end{figure}

\textbf{Vibrational coherent control of the phase transition.} To map out the role of the collective CDW excitation in the transition to the H phase, we perform trARPES measurements with two time-delayed optical pump pulses. The aim is to coherently control the AM (Fig.~\ref{fig:coherent_control}A). The initial pulse excites the AM, whereas the oscillation amplitude is either amplified (in-phase excitation) or suppressed (out-of-phase excitation) by the time-delayed second pulse~\cite{Mihailovic2002_coherentcontrol,Rettig2014_faraday_coherent}. We first demonstrate coherent control of the AM in the C phase in the weak-response regime with both pump pulses at equal fluence far below $F_\mathrm{crit}$ (Supplementary Sec.~\ref{sec:coherent_control}). Utilizing again the transient LHB position at $\overline{\Gamma}$ as a proxy of the CDW symmetry breaking, we observe, upon two-pulse excitation, a strong modulation of the AM oscillation amplitude as a function of pump-pump delay $t_{12}$ with the period of the AM -- indicating effective coherent control. 

Next, we increase the fluence to $F_1=F_2=0.3$~mJ$\cdot$cm$^{-2}$, staying below the critical fluence of the H phase for each individual pulse, but crossing the threshold in combination. By monitoring the LHB at a fixed pump$_1$-probe delay of 25~ps, we probe the system after the new quasi-equilibrium is formed. Varying the pump-pump delay, we track whether the system has been switched to the H phase (strongly reduced LHB intensity) or remains in the C ground state (only minor reduction). Strikingly, the LHB intensity and position feature a pronounced modulation with pump-pump delay, as evinced by the strongly $t_{12}$-dependent EDCs at $\overline{\Gamma}$ (Supplementary Fig.~\ref{fig:S_coherent_switch}). This dependence is highlighted by the phase transition efficiency, a parametrization of the $t_{12}$-dependent LHB intensity (Fig.~\ref{fig:coherent_control}B). The coherent modulation of the switching efficiency with $t_{12}$ at 2.2~THz, close to the AM frequency in the weak-response regime~\cite{demsar2002}, provides direct evidence for the decisive role of the collective CDW excitation in the photoinduced transition and demonstrates exceptional control over the phase transition using a multi-pulse excitation protocol. The distinct dependence of the switching efficiency rules out a temperature-driven transition from C to H phase, since, in a thermal switching scenario, the efficiency would only depend on the total deposited energy and would show no strong dependence on $t_{12}$ within the studied range.

Inspecting the phase of the switching efficiency reveals that its local maxima approximately coincide with constructive amplification of the AM by the second optical pulse, while minima coincide with suppression of the AM by the subsequent pulse (Supplementary Fig.~\ref{fig:S_coherent_phase}). This indicates that, upon constructive amplification of the CDW excitation beyond a certain threshold, the transition to the H phase commences, while suppression impedes switching. Already at a fluence of 0.3~mJ$\cdot$cm$^{-2}$, a faint suppression of the LHB is observed by a single pulse, resulting in a switching efficiency of $\sim 0.5$ for two individual pulses with large pump-pump delay (grey line in Fig.~\ref{fig:coherent_control}B). Remarkably, within the first oscillation cycle, coherent control of the AM allows suppressing the switching efficiency below the level of two individual pulses, demonstrating an effective suppression of the phase transition by the pulse sequence. Coherent amplification and suppression pathways within transient energy surfaces are sketched in Fig.~\ref{fig:coherent_control}, C and D. Note that the one-dimensional energy landscape only serves as a schematic illustration. The offset of the local extrema of the switching efficiency from optimal amplification and suppression of the AM in the low-fluence regime by few 10~fs suggests that the transition is governed by a more complex multi-dimensional energy landscape defined by intra- and interlayer CDW order (Supplementary Fig.~\ref{fig:S_coherent_phase}).

\begin{figure}[!ht]
\centering
\includegraphics[width=\textwidth]{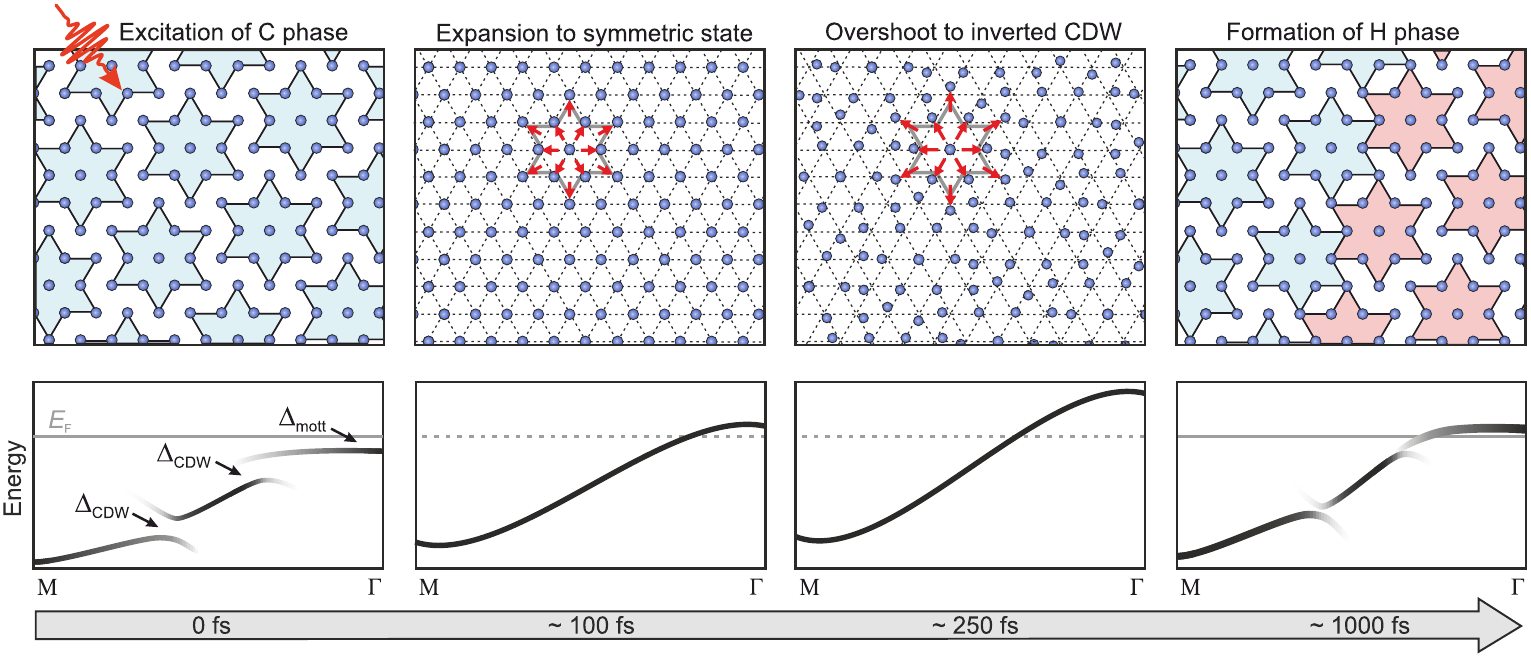}
\caption{\textbf{Illustration of the transition to the H state.} Top: Illustration of the intralayer lattice structure during the phase transition from C to H phase following a collective CDW excitation (see text). Bottom: Corresponding schematic dispersion of the Ta 5$d$ band. The grey horizontal line marks the equilibrium $E_\mathrm{F}$. In the highly out-of-equilibrium state after photoexcitation, the position of $E_\mathrm{F}$ is not well-defined, indicated by the dashed line.}
\label{fig:schematic}
\end{figure}

\section*{Discussion}
Our results demonstrate the ultrafast timescale and vibrational coherent control of the transition from C to H phase and expose the collective CDW excitation as an integral part of the optical switching mechanism. The colossal upwards band shift of the LHB in the strong-response regime, the rapid transition within a few hundred fs, and the control of the switching efficiency through the AM imply a nonthermal, cooperative transition pathway dictated by a transient energy landscape. Based on the observed band-structure dynamics, we propose a microscopic pathway for the photoinduced phase transition (Fig.~\ref{fig:schematic}): Initially, the system is in the Mott-insulating polaronic C phase. The optical pump pulse is absorbed by the electronic system, which thermalizes within tens of fs, reaching a maximum temperature far beyond 1000~K~\cite{perfetti2006, Stojchevska2014Apr} and melting the Mott energy gap. Simultaneously, the strong electronic perturbation displaces the ionic coordinates from their CDW ground-state positions toward the high-temperature symmetric positions, launching a delayed outwards motion towards these new quasi-equilibrium coordinates. Concomitantly, the commensurate CDW energy gaps close and the LHB is renormalized toward higher energies. Intriguingly, strong optical perturbation can lead to an overshoot of the ionic displacement beyond the undistorted high-symmetry positions. While for most CDW materials such an overshoot leads to a CDW pattern that is geometrically equivalent to the initial CDW~\cite{huber2014,Maklar2021_NatCom}, over-driving the star-shaped polaron clusters leads to the formation of a new (inverted) CDW geometry. In the compound 1$T$-TaSe$_2$, which features a similar star-shaped CDW, the formation of an inverted CDW state -- shifting the lattice and charge density from the cluster center to the peripheral -- was observed, associated with a pronounced upwards shift of the low-energy band at $\overline{\Gamma}$~\cite{Zhang2022_CDWinversion}. Similarly in TaS$_2$, the observed colossal shift of the LHB position within 250~fs suggests a CDW overshoot to an inverted pattern. Subsequently, on a timescale of a few hundred fs, the LHB position relaxes, which we assign to the reemergence of polaron clusters. However, the persistent collapse of the Mott-insulating state indicates a transition to the H phase, in which the formation of a network of irregular domain boundaries in each TaS$_2$ layer breaks the interlayer polaron stacking order, quenching the delicate Mott gap of the C phase.

The question arises whether the formation of the H phase plays out through an incoherent formation process or a collective pathway connected to the inverted CDW pattern. The ultrafast switching time suggests the latter, likely facilitated by strong coupling of the over-driven CDW state to other vibrational modes~\cite{Zijlstra_PRB2006,Sciaini2009_Nature} that enable efficient reformation of polaron clusters. Interestingly, as the CDW order in 1$T$-TaS$_2$ is rotated by $\sim13$° with respect to the atomic lattice, previous studies showed that optical excitation of the room-temperature NC phase, which transiently completely suppresses the NC order, leads to the reformation of the NC phase in patches of two degenerate (clockwise and counter-clockwise rotated) CDW domains~\cite{Zong2018_SciAdv}. However, in the H phase, only a single rotational orientation similar to the initial CDW rotation in the C phase is observed~\cite{gerasimenko_2019_Hphase_order}, suggesting that a nonthermal transition pathway, which preserves the orientation, connects C and H phases. Yet, the large number of irregular domain walls also indicates a high degree of disorder during the recovery. The formation of domain walls is attributed to an effective charge injection by the photoexcitation~\cite{Stojchevska2014Apr,Mihailovic2021_APL_memory_devices}. During the recovery of order, the additional charges are accommodated in domain walls within the commensurate structure, effectively increasing the polaron density as neighboring stars within the domain walls share corners~\cite{gerasimenko_2019_Hphase_order}. Local fluctuations and aggregation of excess charges may thus interplay with a cooperative recovery of CDW order, resulting in the formation of the H phase. Additional studies of the ultrafast lattice dynamics are required to elucidate the microscopic details during the reformation of short-range CDW order.

Intriguingly, the H phase can also be induced by optical or voltage pulses with pulse lengths that significantly exceed the period of the AM~\cite{Stojchevska2014Apr,Mihailovic2021_APL_memory_devices}, precluding any collective CDW response. However, the switching by an incoherent charge-carrier injection by ps to ns pulses requires drastically higher switching energy densities. This demonstrates that, while the H phase can also be stabilized by an incoherent excitation, the collective switching pathway enables a more rapid and efficient transition.

\section*{Conclusion}
Using time- and angle-resolved photoemission spectroscopy, we provide a detailed characterization of the electronic band structure and switching dynamics of the metastable hidden state of 1$T$-TaS$_2$ upon femtosecond optical excitation, thereby revealing the key role of a collective charge-density-wave excitation in the nonthermal transition pathway. The exceptional speed and high degree of control over the phase transition in combination with the remarkable stability of the hidden phase underline its unique functionality for memory device applications~\cite{Mihailovic2021_APL_memory_devices}.
Furthermore, we envision that collective excitations of complex symmetry-broken ground states may reveal hidden orders in a variety of materials~\cite{Yoshikawa2021Aug}. Particularly resonant (multi-pulse) excitations of coherences, minimizing detrimental heating of the electrons and lattice, represent a promising route to realizing and understanding new emergent phases in quantum materials.

\section*{Acknowledgements}
We thank P. Sutar (Jožef Stefan Institute) for providing the samples. We thank T. Ritschel (Technische Universität Dresden), M.A. Sentef, D.B. Shin (MPI for the Structure and Dynamics of Matter), M. Müller and L. Jiajun (Paul Scherrer Institut) for enlightening discussions. 
\textbf{Funding:} This work was funded by the Max Planck Society, the European Research Council (ERC) under the European Union's Horizon 2020 research and innovation program (Grant No. ERC-2015-CoG-682843 and OPTOlogic 899794), and the German Research Foundation (DFG) under the Emmy Noether program (Grant No. RE 3977/1), and the DFG research unit FOR 1700. T.P. acknowledges financial support from the Alexander von Humboldt Foundation. The work by P.S.K., J.A.S., S.L.Y., D.L., R.G.M.~and Z.-X.S.~at Stanford and SLAC was supported by the Department of Energy, Office of Basic Energy Sciences, Division of Materials Science and Engineering. D.L.~acknowledges support from the Swiss National Science Foundation, under Fellowship No.~P300P2\_151328. S.-L.Y.~acknowledges support from the Stanford Graduate Fellowship. 
\textbf{Author contributions:} L.R., D.M., and J.M. conceived the experiment; J.M., S.D., J.S., T.P., S.B., and L.R. carried out the XUV trARPES experiments; P.S.K., J.A.S., S.-L.Y., D.L., and R.G.M. carried out the 6eV-laser-ARPES experiments; Y.A.G. conducted and analyzed the STM experiments; J.M. analyzed the trARPES data with support from L.R and J.S.; J.M. wrote the manuscript with support from L.R., Y.A.G., and D.M.; L.R., R.E., M.W., and Z.-X.S. provided the experimental infrastructure; all authors commented on the paper. 
\textbf{Correspondence:} Correspondence should be addressed to \href{mailto:rettig@fhi-berlin.mpg.de}{\nolinkurl{rettig@fhi-berlin.mpg.de}}.

\printbibliography

\clearpage

\onecolumn
\appendix

\setcounter{figure}{0}
\renewcommand\thefigure{S\arabic{figure}}  

\section*{Supplementary Information for "Coherent Light Control of a Metastable Hidden Phase"}

\section*{Materials and Methods}
\underline{XUV-trARPES} 

\begin{figure}[!ht]
\centering
\includegraphics[width=\textwidth]{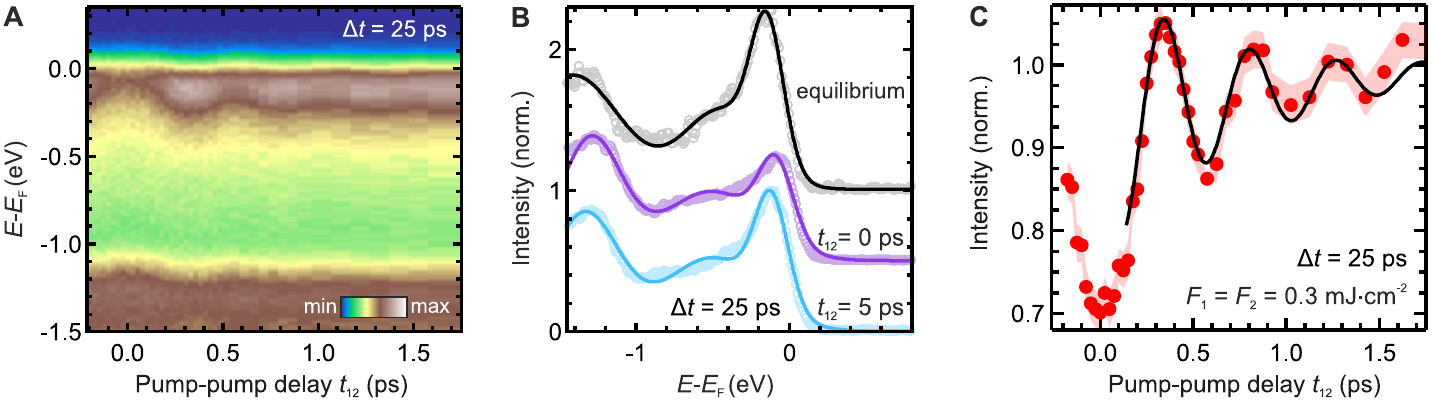}
\caption{\textbf{Coherent control of the H phase using two successive pump pulses.} (\textbf{A}) EDCs at $\overline{\Gamma}$ as function of $t_{12}$ for a fixed pump-probe delay of $\Delta t=25$~ps ($T=160$~K, $F_1=F_2=0.3~$mJ$\cdot$cm$^{-2}$, rep.~rate $=10$~kHz). (\textbf{B}) Exemplary EDCs (circles) with Gaussian fits (solid lines) at $\overline{\Gamma}$ in equilibrium and for selected pump-pump delays $t_{12}$ at $\Delta t=$25~ps ($T=160$~K, $F_1=F_2=0.3$~mJ$\cdot$cm$^{-2}$, rep. rate of 10~kHz, curves vertically offset). (\textbf{C}) LHB peak intensity (red dots) as a function of $t_{12}$ for a fixed pump-probe delay of $\Delta t=25$~ps, extracted from panel \textbf{A}. The intensity is normalized to the value obtained for $t_{12}=5$~ps. The black line marks a damped sinusoidal fit with frequency $f=2.2$~THz.}
\label{fig:S_coherent_switch}
\end{figure}

\begin{figure}[!t]
\centering
\includegraphics[]{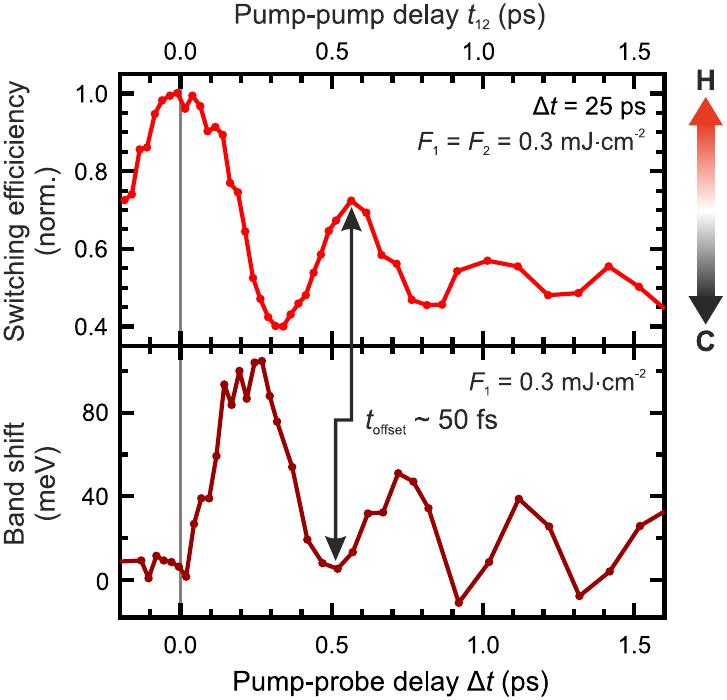}
\caption{\textbf{Comparison of the 2-pulse switching efficiency and the single-pulse AM band oscillation.} Top: Switching efficiency as function of pump-pump delay $t_{12}$ (as presented in Fig.~\ref{fig:coherent_control}) at fixed $\Delta t$ following 2-pulse excitation. Bottom: LHB shift as function of pump-probe delay $\Delta t$ after single-pulse excitation (as presented in Fig.~\ref{fig:pump_probe}). The two curves show approximately inverse behavior, with an offset of only few ten fs between respective maxima/minima. The direct comparison of the two curves shows that 
approximately in-phase excitation (constructive amplification of AM; arrival of second pulse when band shift is at local minimum) leads to efficient switching to the H phase (maximum of switching efficiency). In contrast, when the second pulse arrives approximately out-of-phase (suppression of AM; arrival of second pulse when band shift is at local maximum), switching is inhibited.}
\label{fig:S_coherent_phase}
\end{figure}

Crystals of 1$T$-TaS$_2$ were grown by chemical transport method with iodine as a transport agent. The ARPES measurements were performed in ultra-high vacuum $<1$ $\times$ 10$^{-10}$\,mbar (samples cleaved in-situ), using a table-top high-harmonic-generation trARPES setup~\cite{Puppin2019,Maklar2020_quantitative} ($h\nu_{\mathrm{probe}}$=21.7\,eV, $h\nu_{\mathrm{pump}}$=1.55\,eV, $\Delta E\approx$ 150\,meV, $\Delta t\approx$ 35\,fs) with a pulse picker to vary the repetition rate (from single pulses up to 500~kHz). The samples are positioned with a 6-axis manipulator with cryogenic temperature control (SPECS Carving). Photoelectrons are collected either by a SPECS Phoibos 150 hemispherical analyzer or a SPECS METIS 1000 time-of-flight momentum microscope. The momentum microscope allows parallel acquisition of the 3D photoelectron distribution across a large energy- and in-plane momentum range, advantageous for static band-structure mapping (Figs.~\ref{fig:static}A to E, \ref{fig:S_CDW_gap}, \ref{fig:S_switching_recipe}C and D, \ref{fig:S_temp_ramp}), while the hemispherical analyzer allows for fast data acquisition of 2D energy-momentum cuts, ideal for time-resolved measurements within a limited energy-momentum range at low repetition rates (Figs.~\ref{fig:pump_probe}, \ref{fig:coherent_control}, \ref{fig:S_coherent_switch}, \ref{fig:S_coherent_phase}, \ref{fig:S_coherent}, \ref{fig:S_switching_recipe}A and B).

The EDCs extracted from momentum-microscopy data were integrated over a momentum area of $0.2 \times 0.2$~\AA$^{-2}$. The EDCs extracted from hemispherical-analyzer data were integrated along the momentum-resolved direction over $0.35$~\AA$^{-1}$. For the applied probe energy of 21.7~eV, an inner potential $V_0=20$~eV and work function $\Psi=4.5$~eV of 1$T$-TaS$_2$~\cite{Ngankeu2017_PRB}, we estimate that, within the 3D Brillouin zone, we map the band structure approximately at the Brillouin-zone center with respect to the out-of-plane $k_z$ direction.

The pump and probe spot sizes (FWHM) are $\approx 150 \times 150$~\textmu m$^2$ and $\approx 70 \times 60$~\textmu m$^2$, respectively. All discussed fluences refer to the absorbed fluence, estimated using a complex refractive index of $n$=3.2 and $k$=2.9 at $\lambda$=800~nm. Temporal pump-probe overlap ($\Delta t=0$~fs) was determined from the transient pump-laser-induced population of high-energy states ($E>E_\mathrm{F}+0.5$~eV).

Switching to the H phase has been achieved using optical pulses at 800~nm and an absorbed fluence $> F_\mathrm{crit}\approx0.5$~mJ$\cdot$cm$^{-2}$ at a pulse length of $t_\mathrm{FWHM}=35$~fs.\\

\noindent
\underline{UV-trARPES} 

\begin{figure}[!ht]
\centering
\includegraphics[width=\textwidth]{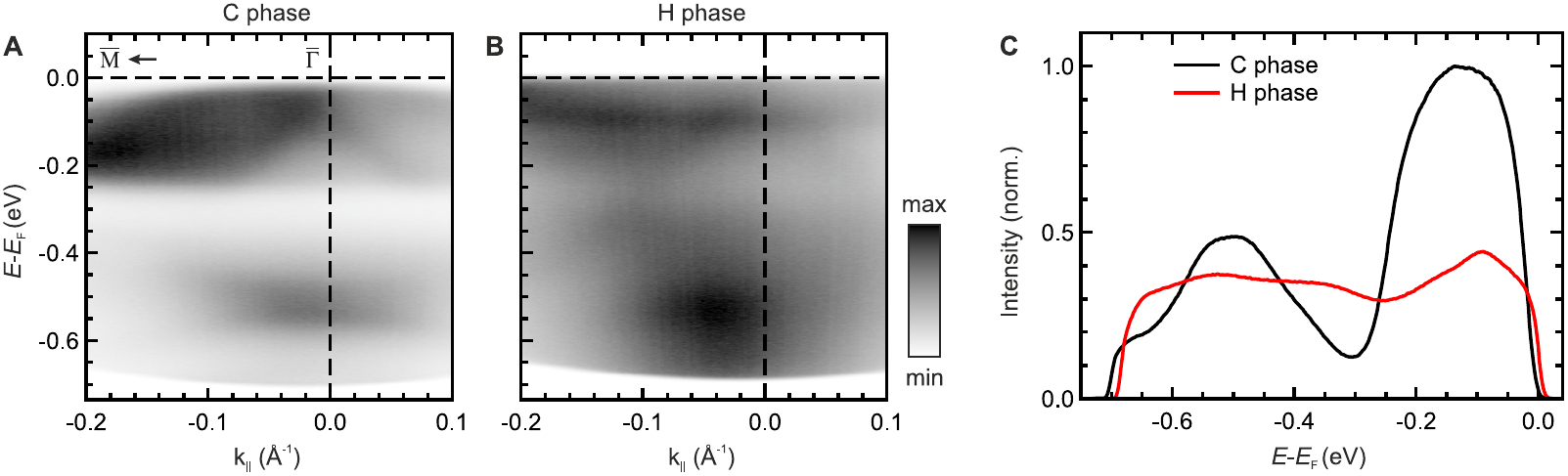}
\caption{\textbf{Static high-energy-resolution ARPES measurements.} (\textbf{A}) Energy-momentum distribution of the pristine C phase and (\textbf{B}) of the photoinduced H phase at a sample temperature of $T=11$~K obtained using a 6~eV laser-based ARPES setup (see methods). (\textbf{C}) EDCs of the respective phases at $\overline{\Gamma}$.}
\label{fig:S_6eV}
\end{figure}

Additional static measurements (Figs.~\ref{fig:static}F, \ref{fig:S_6eV}) with increased energy resolution ($\Delta E<22$~meV) were obtained using a 6~eV-laser-ARPES setup with a Scienta R4000 hemispherical analyzer~\cite{Gauthier2020_JApplPhys}. For the 6eV-laser-ARPES experiments, we used a regenerative amplifier operating at 10~kHz to switch the sample with 10~ms long bursts consisting of 100 p-polarized 830~nm pulses at an absorbed fluence of 1~mJ$\cdot$cm$^{-2}$. The pump beam profile (FWHM) was $363 \times 390$ ~\textmu m$^2$. 6eV-laser-ARPES was acquired with the frequency-quadrupled output of an oscillator operating at 80~MHz repetition rate. The probe beam profile (FWHM) was $41 \times 35$ ~\textmu m$^2$. During the 6-eV ARPES experiments, the sample was kept at 11~K.

\clearpage

\section*{Supplementary Figures}
\label{sec:figures}

\begin{figure}[!ht]
\centering
\includegraphics[width=\textwidth]{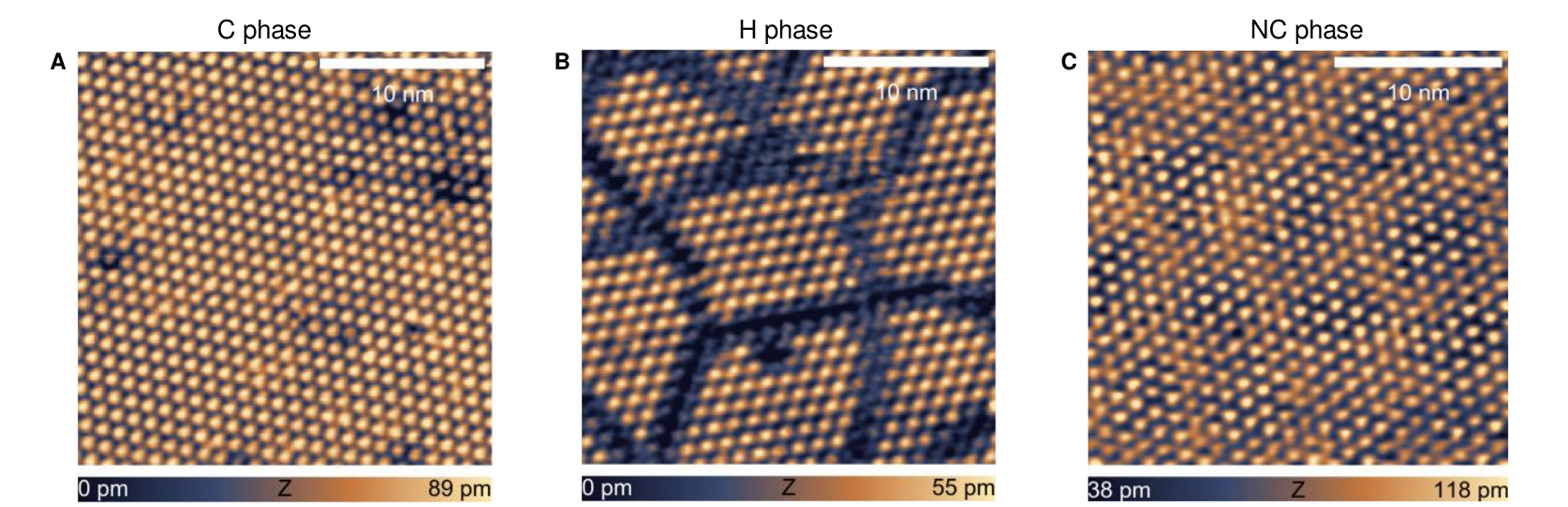}
\caption{\textbf{STM images of the C, H and NC phases of 1$T$-TaS$_2$.} (\textbf{A} and \textbf{B}) The images are taken at $V = -800$~mV at 5~K  and (\textbf{C}) at 300~K. Switching from C to H phase was performed in-situ at 5~K with a single optical pulse at the fluence of 1~mJ$\cdot$cm$^{-2}$, 800~nm central wavelength and 60~fs duration.}
\label{fig:S_stm_images}
\end{figure}

\section*{Supplementary Text}
\section{Extracting Energy Position and Peak Intensity of the LHB}
\label{sec:fitting}
To quantify the (dynamic) energy shift and intensity change of the LHB, we employ a phenomenological description of the experimental EDCs at $\overline{\Gamma}$ using a sum of three Gaussian functions, which yields an excellent description of the data (Fig.~\ref{fig:pump_probe}, B and C). For the fit of the time-resolved data, the FWHM and the position of the central Gaussian located at $E-E_\mathrm{F}\approx-0.7$~eV are kept fixed at the pre-excitation value, while its peak intensity and the remaining parameters of the Gaussian functions describing the LHB at -0.15~eV and the valence band at -1.3~eV are kept as free fit parameters. This fit procedure has been used to extract the modulation of the energetic position and peak intensity of the LHB from time-resolved pump-probe and coherent control data, and has also been applied to evaluate the static temperature ramp (Supplementary Fig.~\ref{fig:S_temp_ramp}).

\begin{figure}[!htb]
\centering
\includegraphics[width=\textwidth]{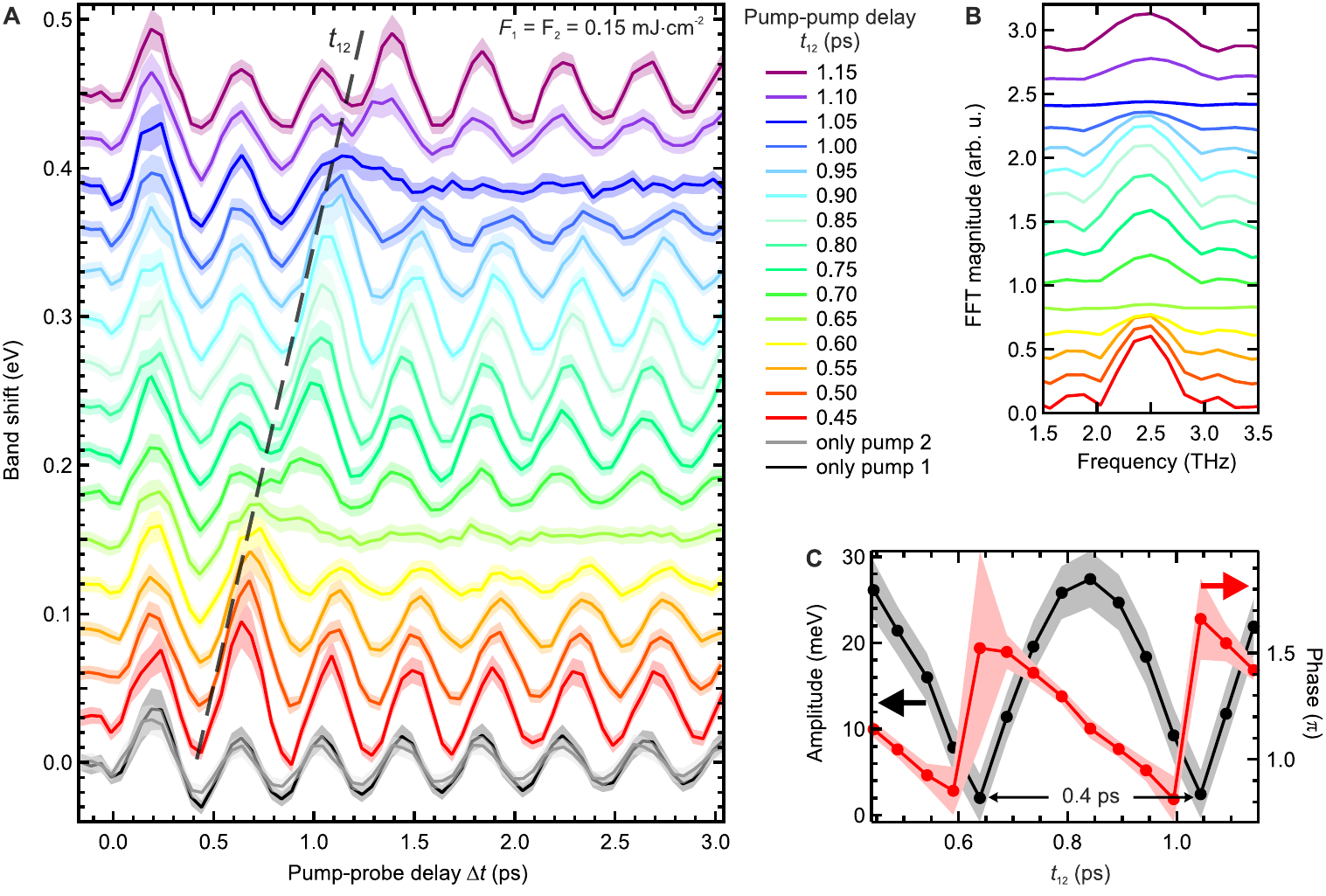}
\caption{\textbf{Coherent control of the CDW AM using two optical pump pulses.} (\textbf{A}) Band shift (solid lines) of the LHB as a function of pump-probe delay $\Delta t$ for various pump-pump delays $t_{12}$ for fluences $F_1=F_2=0.15$~mJ$\cdot$cm$^{-2}$ far below the switching threshold of the H phase (rep. rate 500~kHz, $T=20$~K, curves vertically offset for clarity). The dashed black line indicates the arrival time of the second pump pulse. The black and grey curves mark the reference band shifts induced by the individual pump pulses. The shaded areas correspond to an uncertainty of one standard deviation resulting from the Gaussian fit. (\textbf{B}) Fast Fourier transform magnitude of the band shifts in panel \textbf{A} for $\Delta t>1.5$~ps. (\textbf{C}) Oscillation amplitude (black, left axis) and phase (red, right axis) extracted from sinusoidal fits of the band shifts in \textbf{A} as a function of pump-pump delay $t_{12}$. The shaded areas mark an uncertainty of one standard deviation resulting from the sinusoidal fits.}
\label{fig:S_coherent}
\end{figure}

\section{Coherent Control of the CDW Amplitude Mode in the Weak-response Regime}
\label{sec:coherent_control}


Using trARPES, we demonstrate coherent control of the CDW AM of 1$T$-TaS$_2$ in the C phase in the low-fluence (weak-response) regime. The AM excitation leads to a transient modulation of the energetic position of several bands near the Fermi level. Particularly the LHB energy at $\overline{\Gamma}$ undergoes a very pronounced transient oscillation. Therefore, we track the AM by extracting the LHB energy position at $\overline{\Gamma}$, employing the Gaussian fitting procedure discussed above. Upon excitation by a single pump pulse (see grey and black curves in Supplementary Fig.~\ref{fig:S_coherent}A), a sinusoidal oscillation of the LHB position at a frequency of $\sim$2.4~THz is launched. When exciting the AM by a second pulse (see colored curves), the initial oscillation is either amplified (in-phase excitation) or suppressed (out-of-phase excitation), depending on the pump-pump delay $t_{12}$. This is captured by the Fourier transform magnitudes for various pump-pump delays (Supplementary Fig.~\ref{fig:S_coherent}B), featuring pronounced peaks at 2.4~THz for in-phase excitation ($t_{12}\sim n / 2.4$~THz, with positive integer $n$) and a flat curve for out-of-phase excitation ($t_{12}\sim (n+0.5) / 2.4$~THz). Lastly, we also fit the band shift using a sine function to extract the oscillation amplitude and phase as a function of $t_{12}$ (Supplementary Fig.~\ref{fig:S_coherent}C). The periodicity of amplitude and phase directly reflects the frequency of the AM. 

\section{Parametrization of the Switching Efficiency}
\label{sec:coherent_switching}

To parametrize the phase transition efficiency from C to H phase upon excitation by two successive optical pulses, we fit the $t_{12}$-dependent EDCs at $\overline{\Gamma}$ (Supplementary Fig.~\ref{fig:S_coherent_switch}A) using Gaussian functions as discussed above (exemplary EDCs shown in Supplementary Fig.~\ref{fig:S_coherent_switch}B). From the Gaussian fits, we extract the peak intensity of the LHB as a function of pump-pump delay (Supplementary Fig.~\ref{fig:S_coherent_switch}C). For the parametrization of the switching efficiency (Fig.~\ref{fig:coherent_control}B), the LHB peak intensity at $t_{12}\sim0$~ps is defined as an efficiency of 1 (full switching, H phase), whereas the LHB intensity in the static C ground state is defined as 0 (no switching, C phase). Intensity values between the two extrema are parametrized accordingly to intermediate (partial) switching efficiencies. 

\section{Partial Suppression of CDW gaps in the Hidden Phase}
\label{sec:CDW_gaps}

\begin{figure}[!ht]
\centering
\includegraphics[width=\textwidth]{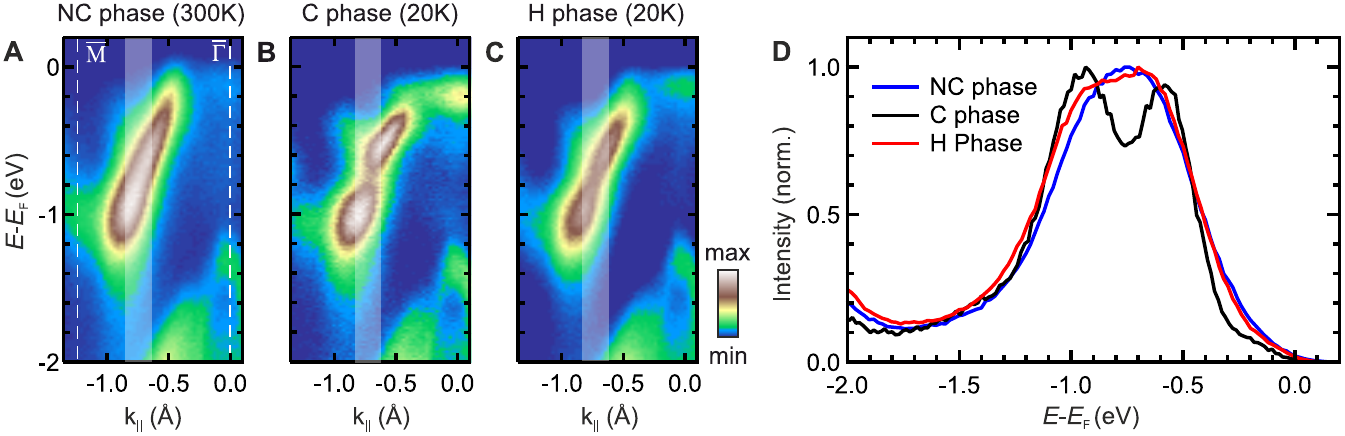}
\caption{\textbf{CDW energy gap.} (\textbf{A} to \textbf{C}) Static energy-momentum cuts along the $\overline{M} - \overline{\Gamma}$ direction for NC, C and H phases. (\textbf{D}) EDCs of the shaded momentum regions in panels \textbf{A} to \textbf{C}, highlighting a partial suppression of the CDW energy gap in the H phase.}
\label{fig:S_CDW_gap}
\end{figure}

In addition to the modification of the LHB, we observe a partial suppression of the commensurate CDW gaps in the H phase (Supplementary Fig.~\ref{fig:S_CDW_gap}). The partial CDW-gap suppression is likely connected to the breaking of interlayer dimerization, as density functional theory calculations predict a strong dependence of the CDW gaps on interlayer stacking order~\cite{Ritschel2015Apr,Lee_2019_PRL_DFT_bandinsulator}. In the H phase, averaging over a large number of microscopic domains with random interlayer arrangements may lead to a blurring of individual, energetically displaced CDW gaps, while also the absolute size of the gaps may be reduced as compared to the bilayer-stacked C phase~\cite{Cho2016_electric_switch_interlayer_stacking}.

\section{Details of the Optical Switching}
\label{sec:switching_details}

\begin{figure}[!thb]
\centering
\includegraphics[scale=1]{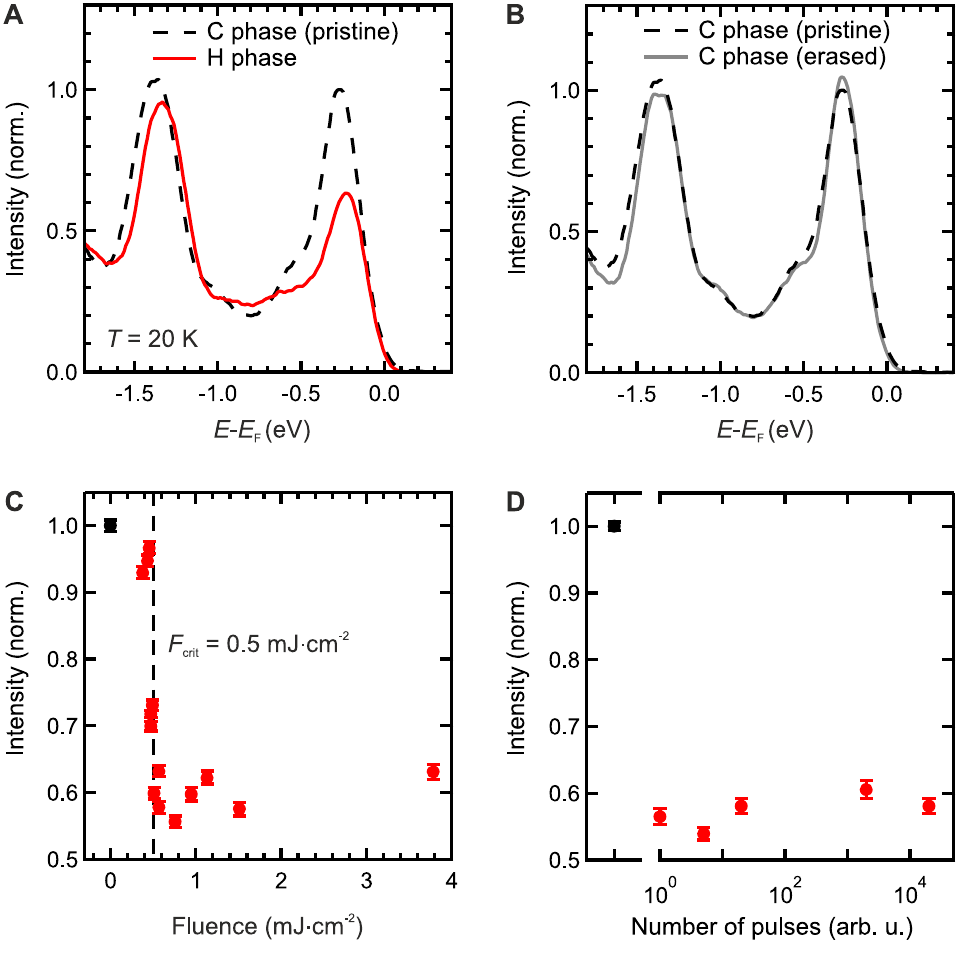}
\caption{\textbf{Characterization of optical switching.} (\textbf{A}) EDCs at $\overline{\Gamma}$ at 20~K of the pristine C phase and after optically inducing the H phase using few optical pulses at $F=2.5$~mJ$\cdot$cm$^{-2}$. (\textbf{B}) EDCs at $\overline{\Gamma}$ of the pristine C phase and after erasing the H phase by laser-based average heating (exposure to the pump laser for 1~s at 500~kHz and $F=0.3$~mJ$\cdot$cm$^{-2}$). (\textbf{C}) Peak intensity of the LHB extracted from Gaussian fits as a function of pump fluence (exposure for 1~s at 20~Hz). The dashed line marks the critical threshold fluence of $F_\mathrm{crit}=0.5$~mJ$\cdot$cm$^{-2}$. After each individual measurement, an optical erase has been performed to revert the system to the C ground state. (\textbf{D}) Intensity of the LHB as a function of number of pump pulses applied to the sample ($F=3.5$~mJ$\cdot$cm$^{-2}$). In panels \textbf{C} to \textbf{D}, the error bars correspond to one standard deviation resulting from the Gaussian fit as described above. The black data points mark the peak intensities of the pristine C phase.}
\label{fig:S_switching_recipe}
\end{figure}

Optical switching from C to H phase and optical erasing, i.e., melting the metastable H phase by laser-heating to restore the C ground state, are highly reproducible over many switching cycles. During the optical erasing procedure, the previously switched sample area is exposed to the pump laser at a repetition rate of 500~kHz at a fluence of a few hundred ~\textmu J$\cdot$cm$^{-2}$, resulting in strong average heating, thus reducing the lifetime of the H phase~\cite{Stojchevska2014Apr}. Exemplary spectra of the pristine C phase, photoinduced H phase, and C phase after optical erasure reveal that the ground state is fully restored by the erasing procedure (Supplementary Fig.~\ref{fig:S_switching_recipe}, A and B). 

For a pump-pulse length of FWHM=35~fs, we find a critical fluence of $F_\mathrm{crit}=0.5$~mJ$\cdot$cm$^{-2}$ at $h\nu=1.55$~eV, as determined from the LHB peak intensity (Supplementary Fig.~\ref{fig:S_switching_recipe}C). Interestingly, we observe a sharp fluence threshold, as the fluence range of intermediate switching between C and H phase amounts to only few ten \textmu J$\cdot$cm$^{-2}$. Note, however, that $F_\mathrm{crit}$ critically depends on the pulse length~\cite{Stojchevska2014Apr}. Furthermore, we find that the full transition to the H phase is induced by a single optical pulse, whereas subsequent pulses show no effect (Supplementary Fig.~\ref{fig:S_switching_recipe}D).

\begin{figure}[!bht]
\centering
\includegraphics[]{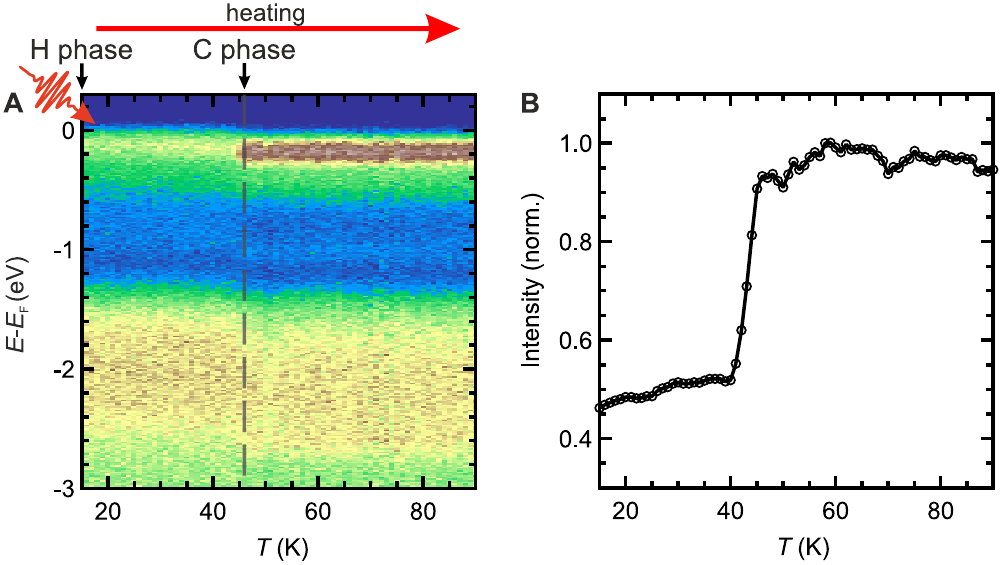}
\caption{\textbf{Stability of the H phase.} (\textbf{A}) EDCs at $\overline{\Gamma}$ with increasing sample temperature $T$ (rate of 1~K per minute) after switching to the H phase at 20~K. (\textbf{B}) Corresponding LHB peak intensity versus $T$.}
\label{fig:S_temp_ramp}
\end{figure}

The critical fluence observed at 20~K during this characterization using the momentum microscope is slightly lower than for the experiments performed at increased temperatures using the hemispherical analyzer (Fig.~\ref{fig:pump_probe}), which may originate from a weak temperature-dependence of the threshold fluence or may stem from a slight inaccuracy of the fluence calibration between different experimental geometries.

The recovery of the metastable H phase is well described by a thermally activated Arrhenius-like behavior~\cite{Vaskivskyi2015Jul}. At 20~K, we observe no recovery of the H phase within the course of 24~hours. However, with increasing temperatures the lifetime strongly decreases, reducing to a few ten seconds at $\sim$50~K~\cite{Vaskivskyi2015Jul}. Thus, heating the sample at a rate of 1~K per minute after optical switching to the H phase at 20~K reveals a relaxation to the C phase at $\sim$45~K, evident from the recovery of the LHB intensity (Supplementary Fig.~\ref{fig:S_temp_ramp}).

\section{Mottness Collapse in the Absence of Intralayer Domain walls}
\label{sec:STM_stacking_order}

Mapping the electronic band structure of the H phase revealed a metallization due to a collapse of the Mott-insulating ground state, which we link to the loss of long-range order along the out-of-plane direction. Here, we present further evidence for the critical role of the stacking order by investigating a buried domain wall in 1$T$-TaS$_2$ using scanning tunneling microscopy (STM), as presented in Supplementary Fig.~\ref{fig:S_stm_interlayer}. After applying a voltage pulse at the position marked by the red cross, we observe the metallization of a patch on the sample surface (indicated by 'hidden') despite the absence of any domain walls in the topmost layer. The STM contrast at the edges of the metallic patch despite the lack of a CDW phase shift indicates buried domain walls in the layer underneath, which further corroborates the strong impact of the stacking order on the low-energy electronic states.

\begin{figure}[!ht]
\centering
\includegraphics[width=\textwidth]{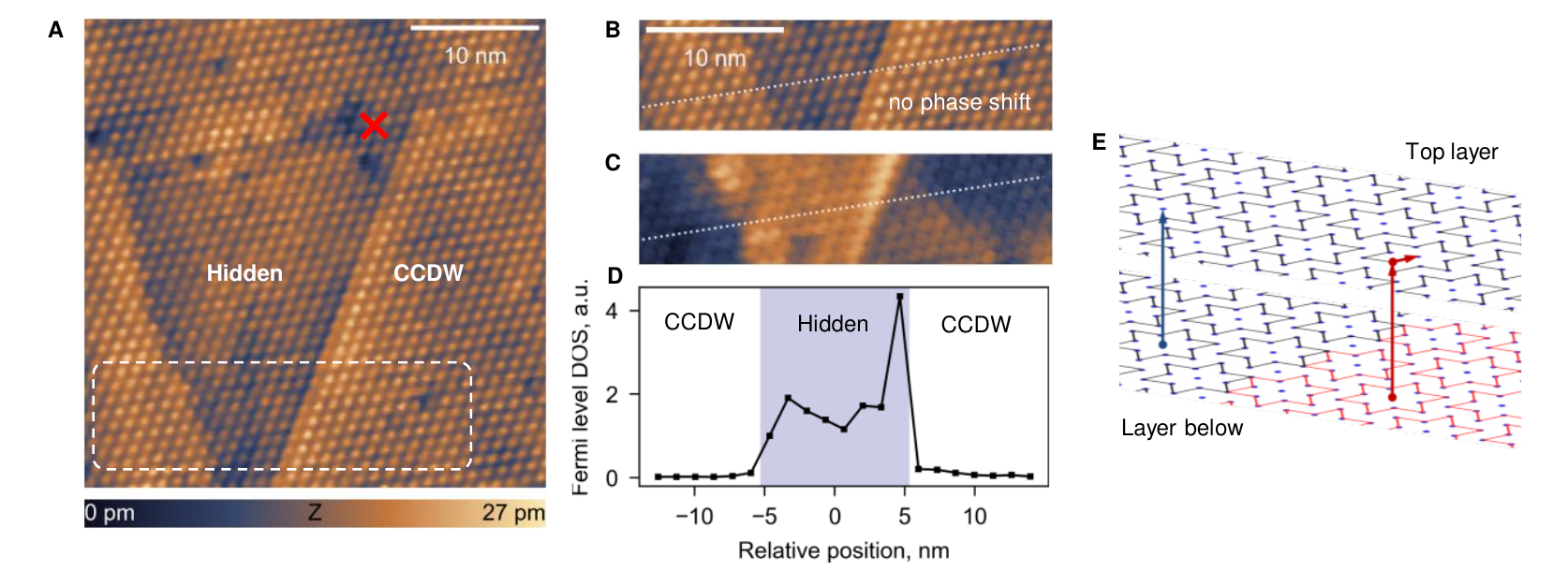}
\caption{\textbf{Mottness collapse in the absence of intralayer domain walls.} (\textbf{A}) Pseudo-topographic STM image of a 1$T$-TaS$_2$ surface ($V = -800$~mV, $I = 1$~nA) after applying a 3~V voltage pulse in the top right corner (red cross).
(\textbf{B}) Close-up inspection of the irregular triangle in \textbf{A} demonstrates the absence of any CDW phase shifts despite the different appearance. (\textbf{C}) In-gap STM image ($V = -50$~mV, $I = 300$~pA) of the area in \textbf{B} demonstrates an enhanced low-energy contrast. Note the bright edge indicating the presence of domain wall states underneath. (\textbf{D}) Spectral weight at the Fermi level extracted from d$I$/d$V$ measurements along the dotted line in \textbf{B} and \textbf{C}, confirming the Mottness collapse in the absence of intralayer domain walls. Note that the in-gap spectral weight is present both at the boundaries and in the middle of the switched area (shaded blue). This confirms the involvement of the coupling between the layers in the Mottness collapse. (\textbf{E}) Speculative arrangement of the Star of David clusters in the top layer and the layer beneath. Red clusters are phase-shifted within the plane with respect to the blue ones. The arrows indicate the inter- and intra-layer shifts respectively.}
\label{fig:S_stm_interlayer}
\end{figure}

\end{document}